\documentclass[preprintnumbers,amsmath,amssymb,nofootinbib,superscriptaddress,floatfix,showkeys,twocolumn]{revtex4-1}
\usepackage[normalem]{ulem}
\usepackage{graphicx}
\usepackage{dcolumn}
\usepackage{bm}
\usepackage{color}
\usepackage[colorlinks,citecolor=blue,urlcolor=blue,linkcolor=blue]{hyperref}
\usepackage{array}
\usepackage{multirow}
\usepackage{enumerate}

\usepackage{subfigure}

\newcommand{\p}{\partial}

\begin{document}
\title{Prospects for improving cosmological parameter estimation with gravitational-wave standard sirens from Taiji}

\author{Ze-Wei Zhao}
\affiliation{Department of Physics, College of Sciences, Northeastern University, Shenyang
110819, China}
\author{Ling-Feng Wang}
\affiliation{Department of Physics, College of Sciences, Northeastern University, Shenyang
110819, China}
\author{Jing-Fei Zhang}
\affiliation{Department of Physics, College of Sciences, Northeastern University, Shenyang
110819, China}
\author{Xin Zhang}
\email{zhangxin@mail.neu.edu.cn}
\affiliation{Department of Physics, College of Sciences, Northeastern University, Shenyang
110819, China}
\affiliation{Ministry of Education's Key Laboratory of Data Analytics and Optimization for Smart Industry, Northeastern University, Shenyang 110819, China}
\affiliation{Center for High Energy Physics, Peking University, Beijing 100080, China}


\begin{abstract}
Taiji, a space-based gravitational-wave observatory, consists of three satellites forming an equilateral triangle with arm length of $3\times 10^6$ km, orbiting around the Sun. Taiji is able to observe the gravitational-wave standard siren events of massive black hole binary (MBHB) merger, which is helpful in probing the expansion of the universe. In this paper, we preliminarily forecast the capability of Taiji for improving cosmological parameter estimation with the gravitational-wave standard siren data. We simulate five-year standard siren data based on three fiducial cosmological models and three models of MBHB's formation and growth. It is found that the standard siren data from Taiji can effectively break the cosmological parameter degeneracies generated by the cosmic microwave background (CMB) anisotropies data, especially for dynamical dark energy models. The constraints on cosmological parameters are significantly improved by the data combination CMB+Taiji, compared to the CMB data alone. Compared to the current optical cosmological observations, Taiji can still provide help in improving the cosmological parameter estimation to some extent. In addition, we consider an ideal scenario to investigate the potential of Taiji on constraining cosmological parameters. We conclude that the standard sirens of MBHB from Taiji will become a powerful cosmological probe in the future.
\end{abstract}


\maketitle


\section{Introduction} \label{sec:intro}

After decades of development, the study of cosmology has entered the era of ``precision cosmology". In order to answer the core questions of the modern cosmology, it is important to precisely measure the cosmological parameters. The observation of cosmic microwave background (CMB) anisotropies from the Planck satellite mission strongly favors a base six-parameter $\Lambda$ cold dark matter ($\Lambda$CDM) model that has been usually viewed as the standard model of cosmology. However, some cracks have recently appeared in the $\Lambda$CDM model. Namely, there appear some tensions between the Planck results (based on the $\Lambda$CDM model) and other low-redshift astrophysical observations, among which the most prominent one is the Hubble constant tension. The Hubble constant tension between the Planck fit value \cite{Aghanim:2018eyx} and the Cepheid-supernova distance ladder measurement value \cite{Riess:2019cxk} is now at about the 4.4$\sigma$ level. Some extended cosmological models considering new physics beyond the standard $\Lambda$CDM cosmology have been proposed to resolve the Hubble tension, but actually no one can truly resolve the tension \cite{Guo:2018ans}.

In fact, the Planck CMB data alone can only measure the base parameters at high precision for the base $\Lambda$CDM model, and when the model is extended to include new parameters the Planck data alone cannot provide precise measurements for them. Therefore, low-redshift cosmological probes, such as the baryon acoustic oscillation (BAO) observation and the type Ia supernova (SN) observation, are needed to combine with the CMB data to break the cosmological parameter degeneracies.

So far, almost all the cosmological probes are based on the electromagnetic (EM) observations. However, recently, the detections of gravitational wave (GW) signals from the compact binary merger events have opened a new window to observe the universe. In particular, the first detection of the GW from the binary neutron star (BNS) merger, i.e. GW170817, initiated the new era of multi-messenger astronomy, which provides a new possibility to explore the universe with the combination of GW and EM observations. GWs can serve as ``standard sirens'' \cite{Schutz1986,Holz:2005df}, because the waveform of GW carries the information of the absolute luminosity distance to the source. If the redshift of the EM counterpart of the GW source can be determined, then a distance-redshift relation can be established and it can be used to probe the expansion history of the universe. The event of GW170817 has been employed to independently measure the Hubble constant, and it has been shown that the result accommodates the results of Planck and SH0ES (Supernovae $H_0$ for the Equation of State) \cite{Abbott:2017xzu}.

A series of recent works indicate that the next-generation ground-based GW detectors can provide lots of standard siren data coming from BNS merger events, which can be used to constrain cosmological parameters at high precision, and thus in the future the standard sirens would be developed into a powerful new cosmological probe \cite{Feeney:2018mkj,Chen:2017rfc,Zhang:2019ylr,Zhang:2018byx,Wang:2018lun,Cai:2016sby,Cai:2017aea,Sathyaprakash:2009xt,Zhao:2010sz,Zhang:2019ple,Zhang:2019loq,Jin:2020hmc,Li:2013lza}. In particular, the GW standard sirens can play a significant role in breaking the cosmological parameter degeneracies generated by the current EM cosmological observations \cite{Wang:2018lun,Zhang:2018byx,Zhang:2019ple,Zhang:2019loq,Jin:2020hmc,Zhang:2019ylr}. Furthermore, space-based GW detectors have also been developed, and the inspiral and merger of massive black hole binary (MBHB) detected by them may offer a unique sample of standard sirens at high redshifts \cite{Tamanini:2016zlh}.

The Laser Interferometer Space Antenna (LISA, http://lisa.nasa.gov/), a space-based GW observatory, aims at detecting GW signals within mHz range (i.e. $10^{-4}$ Hz to 1 Hz). LISA has three satellites forming an equilateral triangle with arm length of $2.5\times 10^6$ km \cite{Audley:2017drz}. This constellation orbits around the Sun at the ecliptic plane behind the Earth.

Similarly, Taiji is a space-based GW observatory proposed by the Chinese Academy of Sciences. Taiji also consists of a triangle of three satellites but with arm length of $3\times 10^6$ km \cite{Wu:2018clg,Hu:2017mde}. Thus, Taiji is more sensitive to low-frequency ranges. Besides, Taiji precedes the Earth in the heliocentric orbit, leading to a LISA-Taiji network, which can significantly improve the measurement precision of sky location \cite{Ruan:2020smc}. Some other studies about Taiji have also been conducted in Refs.~\cite{Wang:2017aqq,Guo:2018npi,Wu:2019thj}. Nevertheless, the forecast for the capability of Taiji in the future cosmological parameter estimation has never been seriously studied. Similar issues concerning LISA and TianQin \cite{Mei:2015joa,Luo:2015ght,Hu:2018yqb,Feng:2019wgq,Wang:2019ryf,Shi:2019hqa} have been investigated in Refs.~\cite{Tamanini:2016zlh,Belgacem:2019pkk,Wang:2019tto}. Therefore, in this paper, we make a forecast for the prospects of Taiji in cosmological parameter estimation by using the simulated standard siren data. We first investigate the capability of constraining cosmological parameters with the standard siren data from Taiji alone. Then we combine the standard siren data from Taiji with other EM cosmological probes to study its effect to break the degeneracies between the cosmological parameters. In addition, we investigate what precision Taiji may achieve if ignoring the redshift error. Throughout this paper, we adopt the units of $c=G=1$.


\section{Methods and data}\label{sec:Method}

\subsection{Configuration of Taiji}
The GW signal from the inspiral of non-spinning MBHB can be modeled by the restricted post-Newtonian (PN) waveform. The strain $h(t)$ measured by a Michelson-type interferometer thus consists of two polarization amplitudes $h_ + (t)$ and $h_\times (t)$,
\begin{equation}
\label{eq:ht}
 h(t) = {F_ + }(t;\theta ,\phi ,\psi ){h_ + }(t) + {F_ \times }(t;\theta ,\phi ,\psi ){h_ \times }(t),
\end{equation}
where $F_{+,\times}$ are the antenna pattern functions, $(\theta,\phi )$ denote the zenith angle and the azimuthal angle of the source relative to the Sun, and $\psi $ is the polarization angle of GW. The antenna pattern functions can be written as
\begin{eqnarray}
 F_{+}^{(1)}(t) &=& \frac{1}{2}\Big({\rm cos}(2\psi)D_{+}(t)-{\rm sin}(2\psi)D_{\times}(t)\Big),  \label{Fplus}\\
 F_{\times}^{(1)}(t) &=& \frac{1}{2}\Big({\rm sin}(2\psi)D_{+}(t)+{\rm cos}(2\psi)D_{\times}(t)\Big)  \label{Fcros}.
\end{eqnarray}

The specific configuration of the GW detector determines the forms of $D_{+,\times}$ that generally depend on the GW frequency. For the inspiral process, we use the low-frequency approximation given in Ref.~\cite{Ruan:2020smc},
\begin{widetext}
\begin{eqnarray}
  D_{+}(t) &=& \frac{\sqrt{3}}{64}\bigg[-36{\rm sin}^2\theta\,{\rm sin}\big(2\alpha(t)-2\beta\big)+\big(3+{\rm cos(2\theta)}\big)
\left.\bigg({\rm cos}(2\phi)\Big(9\sin(2\beta)-{\rm sin}\big(4\alpha(t)-2\beta\big)\Big)+{\rm sin}(2\phi)\right.\nonumber \\
&&\left.\times\Big({\rm cos}\big(4\alpha(t)-2\beta\big)-9\cos(2\beta)\Big)\bigg)\right.-4\sqrt{3}{\rm sin}(2\theta)\Big({\rm sin}\big(3\alpha(t)-2\beta-\phi\big)-3{\rm sin}\big(\alpha(t)-2\beta+\phi\big)\Big)\bigg]  \,,\label{Dplus}  \\
D_{\times}(t) &=& \frac{1}{16}\bigg[\sqrt{3}{\rm cos}\theta\Big(9{\rm cos}(2\phi-2\beta)-{\rm cos}\big(4\alpha(t)-2\beta-2\phi\big)\Big)-6{\rm sin}\theta\Big({\rm cos}\big(3\alpha(t)-2\beta-\phi\big)+3{\rm cos}\big(\alpha(t)\nonumber \\
&&-2\beta+\phi\big)\Big)\bigg]  \,,\label{Dcros}
\end{eqnarray}
\end{widetext}
where $\alpha=2\pi f_m t+\kappa$ is the orbital phase of the guiding center, and $\beta = 2\pi n/3$ ($n=0,1,2$) is the relative phase of three spacecrafts. Here $\kappa$ is the initial ecliptic longitude of the guiding center and $f_m=1/{\rm yr}$.

Taiji can be equivalently considered as a combination of two independent interferometers with an azimuthal difference of $\pi/4$ \cite{Cutler:1997ta}. Thus another equivalent antenna pattern function is
\begin{eqnarray}
F_{+ ,\times}^{(2)}(t;\theta ,\phi ,\psi )=F_{+ ,\times}^{(1)}(t;\theta ,\phi- \frac{\pi}{4} ,\psi ).
\end{eqnarray}
The Fourier transform in frequency-domain of the strain in Eq. \eqref{eq:ht} can be obtained by using the stationary phase approximation,
\begin{eqnarray}
\label{eq:strainf}
\tilde{h}(f)=-\left(\frac{5\pi}{24}\right)^{1/2}M_{\rm c}^{5/6}\left[\frac{(\pi f)^{-7/6}}{D_{{\rm eff}}}\right] e^{-i\Psi}.
\end{eqnarray}
In this paper, ``$\sim$" above a function denotes the Fourier transform of the function. $M_{\rm c} = (1+z)\eta^{3/5} M$ is the observed chirp mass, $M=m_1+m_2$ is the total mass and $\eta=m_1 m_2/M^2$ is the symmetric mass ratio. The definition of the function $\Psi$ can be found in Ref.~\cite{Sathyaprakash:2009xs}. The effective luminosity distance to the source $D_{\rm eff}$ is given by
\begin{equation}
D_{{\rm eff}}=d_{\rm L} \left[F^{2}_{+}\left(\frac{1+{\rm cos}^2 \iota}{2}\right)^2+F^{2}_{\times} {\rm cos}^2 \iota
\right]^{-1/2},
\end{equation}
where $d_{\rm L}$ is the absolute luminosity distance to the source and $\iota$ is the inclination angle between the orbital angular momentum axis of the binary and the line of sight.

To study the signal in the Fourier space, we should denote time in terms of frequency observed on the Earth using the function \cite{Krolak:1995md,Buonanno:2009zt}
\begin{equation}
   t(f) = t_{\rm c} - \frac{5}{256} M_{\rm c} ^{-5/3}(\pi f)^{-8/3},
\end{equation}
where $t_{\rm c}$ is the coalescence time of MBHB. In our analysis, we only consider the leading term \cite{Ruan:2020smc} and set $t_{\rm c}=0$.

The combined signal-to-noise ratio (SNR) for the network of two equivalent independent interferometers is
\begin{equation}
\rho=\sqrt{\sum\limits_{i=1}^{2}(\rho^{(i)})^2},
\label{euqa:rho}
\end{equation}
where $\rho^{(i)}=\sqrt{({h}^{(i)}|{h}^{(i)})}$, with the inner product being defined as
\begin{align}
(a|b)\equiv4\int_{f_{\rm min}}^{f_{\rm max}}\frac{\tilde{a}(f)\tilde{b}^{*}(f)+\tilde{a}^{*}(f)\tilde{b}(f)}{2}\frac{df}{S_{\rm n}(f)},
\end{align}
where $S_{\rm n}(f)$ is the one-sided noise power spectral density. We have limited the integral within $[f_{\rm min},f_{\rm max}]$ for simplicity of calculation. For Taiji, $S_{\rm n}(f)$ is adopted from Ref.~\cite{Guo:2018npi}, the lower and upper cutoff frequencies are chosen to be $f_{\rm min}=10^{-4}\,\rm Hz$ and $f_{\rm max}=c/2\pi L\simeq0.05\frac{\rm Gm}{L}\,\rm Hz$ \cite{Cai:2017aea} with $L$ being the arm length of Taiji, and the SNR threshold is 8 for a detection.

\subsection{Property of GW source}\label{subsec:MBH}

The unclear birth mechanisms of MBHB lead to the uncertainties in predicting the event rate of MBHB. Based on a semi-analytical galaxy formation model, three MBHB models defined by various combinations of mechanism of seeding and delay are presented in Ref.~\cite{Klein:2015hvg}.
\begin{enumerate}[(1)]
\item Model pop III: This model assumes that light MBH seeds from pop III star and there are delays between MBH merger and host galaxy coalescence.
\item Model Q3d: In this model, MBH is assumed to seed from the collapse of protogalactic disks. As in the pop III model, the delays between MBH merger and host galaxy coalescence are also considered.
\item Model Q3nod: This model assumes the same seed of MBH as model Q3d, but it ignores the delays. Compared to the first two realistic and conservative scenarios, this is a relatively optimistic scenario.
\end{enumerate}
\begin{figure}[h]
\begin{center}
\includegraphics[width=\linewidth,angle=0]{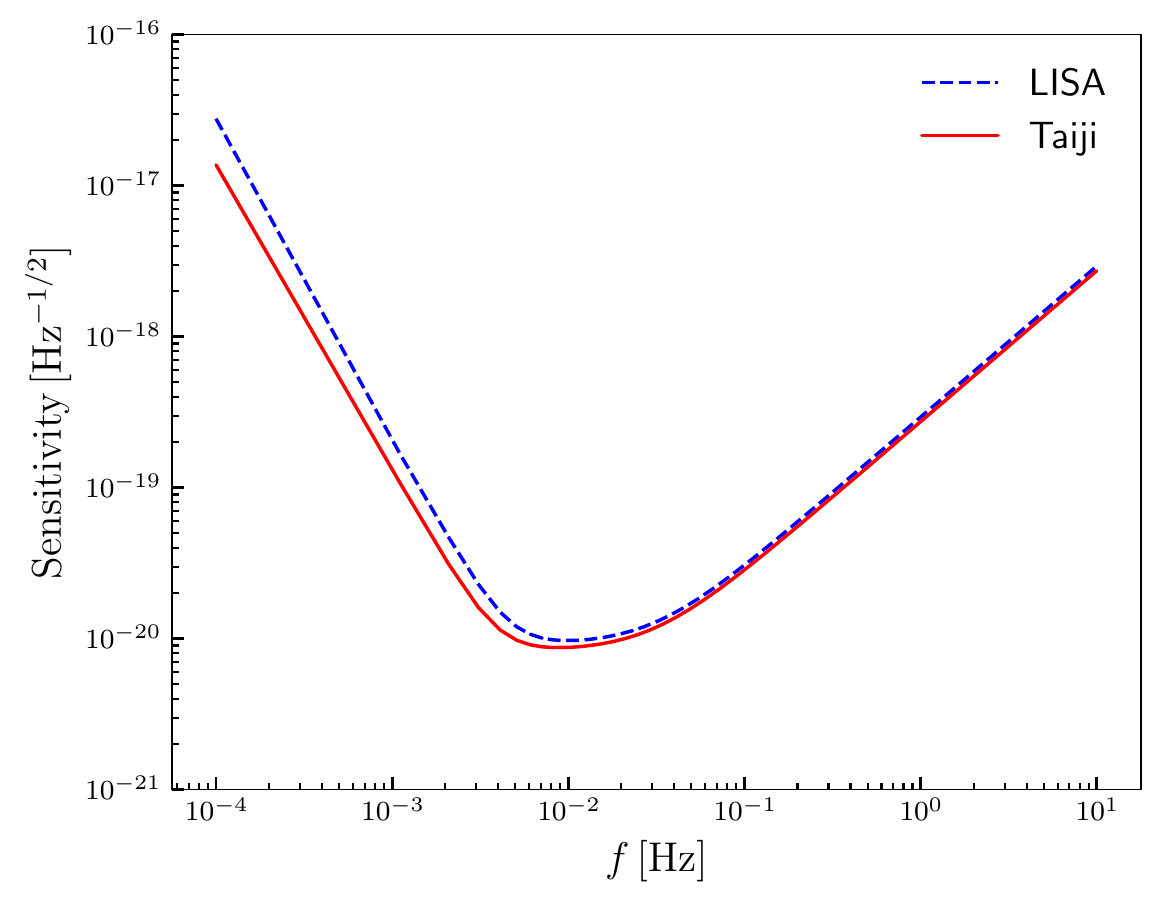}
\end{center}
\caption{The sensitivity curves of Taiji (solid) and LISA with the configuration N2A2M5L6 (dashed). Taiji shows similar behaviour with LISA, with a better performance at low frequencies.} \label{sen}
\end{figure}

Taiji is very similar to the configuration N2A2M5L6 of LISA, as can be seen from their sensitivity curves in Fig.~\ref{sen}. Therefore, we assume that the Taiji's detection rates of MBHBs are identical to those of the LISA. This assumption is reasonable for a preliminary estimate considering the huge uncertainties in MBHB. We select the detection rates and the redshift distribution from Ref.~\cite{Tamanini:2016uin} for a five-year lifetime of Taiji. For the pop III, Q3d, and Q3nod models, we consider 28, 27, and 41 standard siren events, respectively. Without loss of generality, we evenly sample the black hole's mass, the position angles ($\theta$, $\phi$), the polarization angle $\psi$, and the inclination angle $\iota$ in the five parameter intervals: [$10^{4}$, $10^{7}$] $M_\odot$, [0, $\pi$], [0, 2$\pi$], [0, $\pi$], and [0, $\pi$], respectively, where $M_\odot$ is the solar mass.

\subsection{Method of constraining cosmological parameters}

The Fisher matrix can be used to  propagate the errors between different parameters. In this paper, we focus on how the measurement errors of the luminosity distance are propagated to the constraint errors of the cosmological parameters. For a cosmological model with parameters $\theta_i$, the entries of the Fisher matrix are defined as
\begin{align}
F_{ij}=\sum_n \frac{1}{\sigma^2_{d_{\rm L}}(z_n)} \frac{\p d_{\rm L}(z_n)}{\p \theta_i}\bigg|_{\rm fid} \frac{\p d_{\rm L}(z_n)}{\p \theta_j}\bigg|_{\rm fid},
\end{align}
where the sum is over all MBHB merger events, $z_n$ is the redshift of the $n$th GW event, and the derivatives of $d_{\rm L}$ are evaluated at the fiducial values of cosmological models. In order to clearly compare different constraints in the same parameter plane, we set the best-fit values of CMB+BAO+SN as the fiducial values to simulate the standard siren data.

The measurement error of luminosity distance $\sigma_{d_{\rm L}}$ in our work consists of the following aspects.
\begin{enumerate}[(1)]
\item Instrumental error: By applying the Fisher matrix to the GW waveform and assuming that $d_{\rm L}$ is independent of other parameters, the instrumental error on the measurement of luminosity distance is \cite{Tamanini:2016zlh,Feng:2019wgq}
\begin{align}
\sigma_{d_{\rm L}}^{\rm inst}\simeq \frac{d_{\rm L}}{\rho}. \label{instru}
\end{align}

It should be noted that this estimate is relatively optimistic, since there is a strong correlation between $d_{\rm L}$ and $\iota$ in the real analysis \cite{Zhao:2017cbb,Zhao:2017imr,Zhao:2018gwk}. In this work, we assume that $\iota$ could be measured precisely in the future by GW detector networks and EM observations, thus breaking the distance-inclination degeneracy.

\item Weak-lensing error: The main systematic error at high redshift comes from weak lensing. For the fitting formula of weak-lensing error, we adopt the form in Ref.~\cite{Tamanini:2016zlh},
\begin{align}\label{eq:DECIGOlensingError}
\sigma_{d_{\rm L}}^{\rm lens}(z)=d_{\rm L}(z)\times 0.066\bigg[\frac{1-(1+z)^{-0.25}}{0.25}\bigg]^{1.8}.
\end{align}
Notice that we consider a de-lensing factor of two in our simulation for an optimistic forecast.
\item Peculiar velocity error: The error due to the peculiar velocity of the source should also be included \cite{Kocsis:2005vv},
\begin{align}
\sigma_{d_{\rm L}}^{\rm pv}(z)=d_{\rm L}(z)\times\bigg[1+\frac{(1+z)}{H(z)d_{\rm L}(z)}\bigg]\sqrt{\langle v^{2}\rangle},
\end{align}
where $H(z)$ is the Hubble parameter and we roughly set the peculiar velocity of the source with respect to the Hubble flow $\sqrt{\langle v^{2}\rangle}=500\,\mathrm{km\,s^{-1}}$.

\item Redshift error: The process of measuring the redshift of a GW source with optical probes also produces error. This error could be ignored if the redshift is measured spectroscopically, but it should be taken into account when using photometric redshift for the distant source. For the latter, we estimate the error on the redshift as $(\Delta z)_n\simeq 0.03(1+z_n)$ \cite{Ilbert:2013bf} and propagate it to the error on $d_{\rm L}$. As indicated by the analysis in Refs. \cite{Klein:2015hvg,Tamanini:2016zlh}, the flares and jets of MBHBs could be most detected by the Square Kilometre Array (SKA, http://www.skatelescope.org) and the redshifts of sources could be measured by the optical/infrared facilities like the Extremely Large Telescope (ELT, http://www.eso.org/sci/facilities/eelt/). Whether an event is measured spectroscopically or photometrically is determined by the apparent magnitude of host galaxy relative to the threshold of ELT \cite{Davies:2010fv}. Since Taiji is similar to LISA and the redshift measurements mainly depend on EM observations, we assume the numbers of photometric observation events of Taiji to be consistent with those of LISA in Ref.~\cite{Tamanini:2016zlh}. For the pop III, Q3d, and Q3nod models, we consider 17, 14, and 25 photometric observation events, respectively.
\end{enumerate}

Notice that the distributions of the locations and masses of MBHBs adopted in Sec. \ref{subsec:MBH} are a preliminary estimate. In principle, the mass functions for MBHBs may depend on the three MBHB evolution models \cite{Klein:2015hvg}, and both $\cos\theta$ and $\cos\iota$ should distribute evenly from $-1$ to 1 for an isotropic distribution of GW sources \cite{Li:2013lza}. However, for the MBHB standard sirens up to high-redshift, the dominating contributions to the total uncertainties of $d_{\rm L}$ are from the weak-lensing error and the redshift error, whereas the location and mass of an MBHB mainly affect the instrumental error of $d_{\rm L}$. It is expected that the distribution functions of GW sources' masses and locations would not make much difference to our major results.

\begin{figure}[h]
\begin{center}
\includegraphics[width=\linewidth,angle=0]{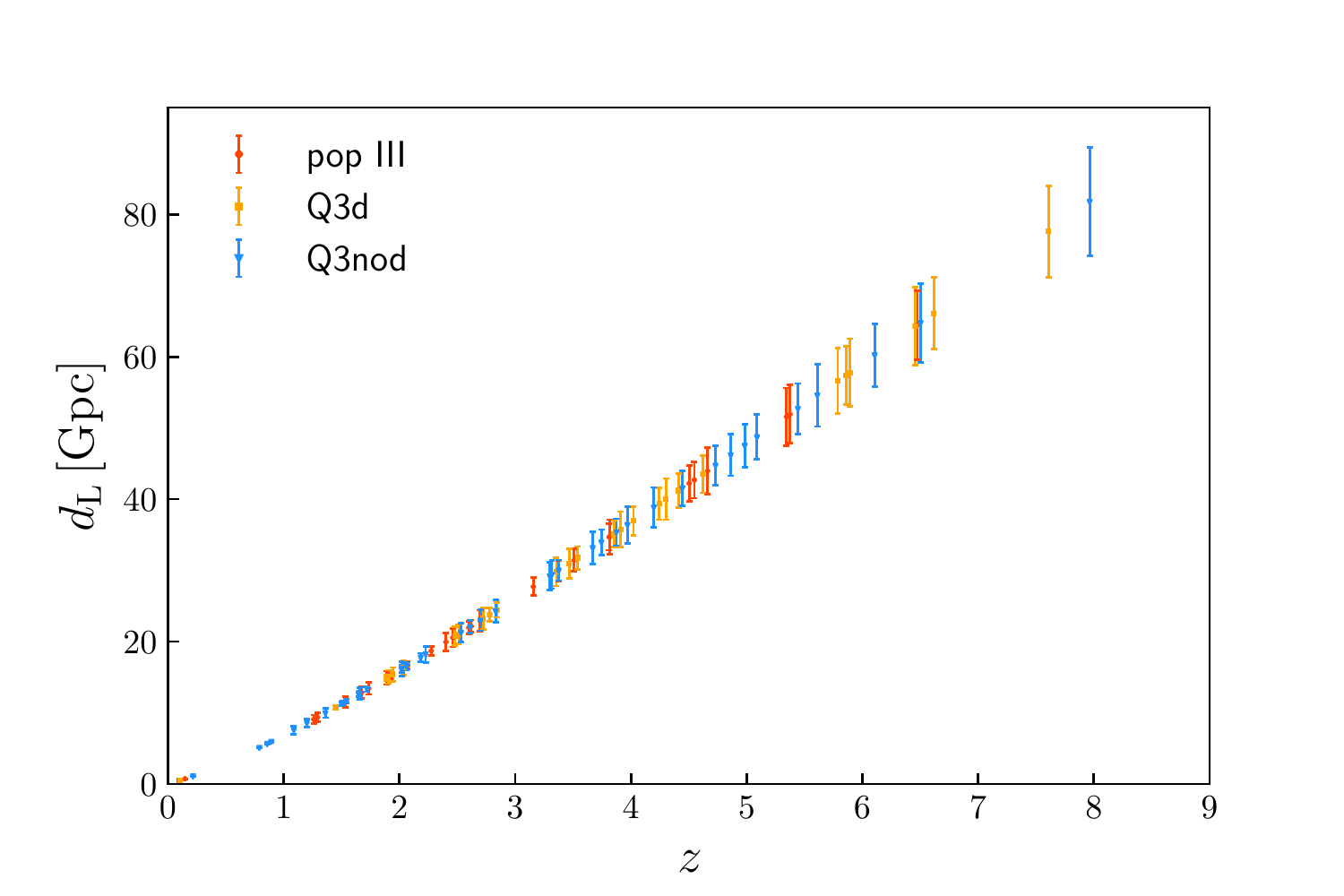}
\end{center}
\caption{The simulated standard siren events of Taiji based on the $\Lambda$CDM model. The redshift distributions  and measurement errors of standard siren events are shown in the figure. Three MBHB models are denoted by different colors, i.e. red (pop III), orange (Q3d), and blue (Q3nod).} \label{sirens}
\end{figure}

To show the redshift distributions and total measurement errors of standard siren events for the three MBHB models, we plot the simulated data of Taiji based on the $\Lambda$CDM model in Fig. \ref{sirens}.

In order to get the necessary statistics, for each selected combination of MBHB model and cosmological model, we construct 1000 catalogs to calculate the mean errors of cosmological parameters using Fisher matrix. The catalog that gives the errors closest to the mean errors is selected as a representative. This representative GW data set is used to infer the posterior probability distributions of cosmological parameters by the Markov-chain Monte Carlo analysis \cite{Lewis:2002ah}. For the GW standard siren measurement with $n$ simulated data points, we can write its $\chi^2$ as
\begin{align}
\chi_{\rm GW}^2=\sum\limits_{n}\left[\frac{\bar{d}_L(z_n)-d_{\rm L}(z_n;\theta_i)}{\bar{\sigma}_{d_{\rm L}}(z_n)}\right]^2,
\label{equa:chi2}
\end{align}
where $\bar{d}_L(z_n)$ and $\bar{\sigma}_{d_{\rm L}}(z_n)$ are the $n$th luminosity distance and the error of luminosity distance, respectively, from the simulated GW data.

%
\subsection{Cosmological models and electromagnetic observational data}\label{subsec:model}

For a dark energy with the equation of state (EoS) $w(z)=p_{\rm de}(z)/\rho_{\rm de}(z)$, the Hubble parameter $H(z)$ can be given by the Friedmann equation,
\begin{align}
H^2(z) = &~H_0^2 \left\{ (1 - {\Omega _m})\exp \left[3\int_0^z {\frac{{1 + w(z')}}{{1 + z'}}} d z'\right] \right.\nonumber\\
           &~+{\Omega _m}{(1 + z)^3}\bigg\},
\label{equa:H}
\end{align}
and the luminosity distance is given by
\begin{align}\label{eq:dL0}
d_{\rm L}(z)=(1+z)\int_{0}^{z}\frac{dz^{\prime}}{H(z^{\prime})}.
\end{align}
Here $H_0=100h\, {\rm km\, s^{-1}\, Mpc^{-1}}$ is the Hubble constant and ${\Omega _m}$ is the current matter density parameter.

In this work, we consider three cosmological models: (i) $\Lambda$CDM model: the standard cosmological model, in which dark energy is described by a cosmological constant $\Lambda$ with $w(z) = -1$; (ii) $w$CDM model: the simplest dynamical dark energy model, in which the EoS of dark energy is fixed to be a constant, i.e. $w(z) = w$; (iii) Chevallier--Polarski--Linder (CPL) model: the parameterized dynamical dark energy model with $w(z)=w_{\rm{0}}+w_{\rm{a}}z/(1+z)$ \cite{Chevallier:2000qy,Linder:2002et}.

For the EM observational data, we consider CMB, BAO, and SN in this work. For the CMB observation, we use the ``Planck distance priors" derived from the Planck 2018 data release \cite{Chen:2018dbv}, instead of the full power spectra data from Planck. For the BAO data, we consider the measurements from 6dFGS at $z_{\rm eff} = 0.106$ \cite{Beutler:2011hx}, SDSS-MGS at $z_{\rm eff} = 0.15$ \cite{Ross:2014qpa}, and BOSS-DR12 at $z_{\rm eff} = 0.38$, 0.51, and 0.61 \cite{Alam:2016hwk}. For the SN data, we use the latest sample from the Pantheon compilation \cite{Scolnic:2017caz}.

\section{Results and discussions} \label{sec:Result}

We consider three fiducial cosmological models described in Sec.~\ref{subsec:model}, and for each of them we consider three MBHB models described in Sec.~\ref{subsec:MBH}. The results of these cases with some relevant discussions are displayed in this section. We first report the constraint results from the standard siren data of Taiji alone, and then combine these data with the EM observations.

\subsection{Standard sirens alone}\label{subsec:st}

\begin{table*}[!htb]
\caption{The 1$\sigma$ errors on the cosmological parameters in the $\Lambda$CDM, $w$CDM, and CPL models, by using the CMB, CMB+BAO, CBS, Taiji, CMB+Taiji, and CBS+Taiji data. Here, CBS stands for CMB+BAO+SN. The three values in a cell in the columns of Taiji, CMB+Taiji, and CBS+Taiji represent pop III, Q3d, and Q3nod, respectively, from top to bottom.}
\label{tab:full}
\begin{center}{\centerline{
\begin{tabular}{|c|c|m{2cm}<{\centering}|m{2cm}<{\centering}|m{2cm}<{\centering}|m{2cm}<{\centering}|m{2cm}<{\centering}|m{2cm}<{\centering}|}
\hline
          model       & parameter   & CMB & CMB+BAO& CBS & Taiji & CMB+Taiji & CBS+Taiji \\ \hline
                      & \text{} &  &   &     & $2.2\times 10^{-2}$ & $6.8\times 10^{-3}$ & $5.3\times 10^{-3}$ \\
\multirow{4}{*}{$\Lambda$CDM} & $\sigma(\Omega_m)$ & $8.5\times 10^{-3}$ & $6.1\times 10^{-3}$ & $6.0\times 10^{-3}$ & $2.4\times 10^{-2}$ & $7.3\times 10^{-3}$ & $5.5\times 10^{-3}$ \\
                      & \text{} &  &   &     & $1.6\times 10^{-2}$ & $6.0\times 10^{-3}$ & $4.9\times 10^{-3}$ \\ \cline{2-8}
                      & \text{} &   &  &     & $9.8\times 10^{-3}$ & $4.8\times 10^{-3}$ & $3.8\times 10^{-3}$ \\
                      & $\sigma(h)$  & $6.1\times 10^{-3}$ & $4.5\times 10^{-3}$ & $4.4\times 10^{-3}$ & $1.2\times 10^{-2}$ & $5.2\times 10^{-3}$ & $4.0\times 10^{-3}$ \\
                      & \text{} &  &   &     & $7.0\times 10^{-3}$ & $4.3\times 10^{-3}$ & $3.5\times 10^{-3}$ \\ \hline
                      & \text{} &   &  &     & $2.3\times 10^{-2}$ & $8.8\times 10^{-3}$ & $5.8\times 10^{-3}$ \\
\multirow{7}{*}{$w$CDM} & $\sigma(\Omega_m)$ & $5.8\times 10^{-2}$ & $1.2\times 10^{-2}$ & $7.9\times 10^{-3}$ & $2.3\times 10^{-2}$ & $8.0\times 10^{-3}$ & $5.7\times 10^{-3}$ \\
                      & \text{} &  &   &     & $2.1\times 10^{-2}$ & $7.1\times 10^{-3}$ & $5.2\times 10^{-3}$ \\ \cline{2-8}
                      & \text{} &   &  &     & $2.2\times 10^{-2}$ & $1.0\times 10^{-2}$ & $6.3\times 10^{-3}$ \\
                      & $\sigma(h)$  & $6.7\times 10^{-2}$ & $1.4\times 10^{-2}$ & $8.4\times 10^{-3}$ & $2.9\times 10^{-2}$ & $8.9\times 10^{-3}$ & $6.2\times 10^{-3}$ \\
                      & \text{} &  &   &     & $1.5\times 10^{-2}$ & $8.0\times 10^{-3}$ & $5.7\times 10^{-3}$ \\ \cline{2-8}
                      & \text{} &   &  &     & $2.4\times 10^{-1}$ & $4.5\times 10^{-2}$ & $3.1\times 10^{-2}$ \\
                      & $\sigma(w)$  & $2.3\times 10^{-1}$ & $5.7\times 10^{-2}$ & $3.4\times 10^{-2}$ & $3.1\times 10^{-1}$ & $4.1\times 10^{-2}$ & $3.1\times 10^{-2}$ \\
                      & \text{} &  &   &     & $1.8\times 10^{-1}$ & $3.8\times 10^{-2}$ & $2.9\times 10^{-2}$ \\ \hline
                      & \text{} &   &  &     & $1.5\times 10^{-1}$ & $2.2\times 10^{-2}$ & $6.9\times 10^{-3}$ \\
                      & $\sigma(\Omega_m)$ & $8.8\times 10^{-2}$ & $2.7\times 10^{-2}$ & $8.2\times 10^{-3}$ & $1.6\times 10^{-1}$ & $3.5\times 10^{-2}$ & $8.0\times 10^{-3}$ \\
                      & \text{} &  &   &     & $1.5\times 10^{-1}$ & $1.7\times 10^{-2}$ & $6.8\times 10^{-3}$ \\ \cline{2-8}
                      & \text{} &   &  &     & $4.0\times 10^{-2}$ & $2.3\times 10^{-2}$ & $6.9\times 10^{-3}$ \\
\multirow{4}{*}{CPL}& $\sigma(h)$  & $9.9\times 10^{-2}$ & $2.5\times 10^{-2}$ & $8.7\times 10^{-3}$ & $4.2\times 10^{-2}$ & $3.9\times 10^{-2}$ & $8.5\times 10^{-3}$ \\
                      & \text{} &  &   &     & $3.1\times 10^{-2}$ & $1.8\times 10^{-2}$ & $6.8\times 10^{-3}$ \\ \cline{2-8}
                      & \text{} &  &   &     & $6.6\times 10^{-1}$ & $3.1\times 10^{-1}$ & $8.4\times 10^{-2}$ \\
                      & $\sigma(w_0)$ & $9.9\times 10^{-1}$ & $2.9\times 10^{-1}$ & $9.1\times 10^{-2}$ & $7.2\times 10^{-1}$ & $3.6\times 10^{-1}$ & $8.8\times 10^{-2}$ \\
                      & \text{} &  &   &     & $4.7\times 10^{-1}$ & $2.3\times 10^{-1}$ & $7.9\times 10^{-2}$ \\ \cline{2-8}
                      & \text{} &  &   &     & $ 3.2 $ & $8.8\times 10^{-1}$ & $2.7\times 10^{-1}$ \\
                      & $\sigma(w_a)$ & 3.2 & $7.7\times 10^{-1}$ & $3.3\times 10^{-1}$ & $3.3$ & $1.0$ & $3.1\times 10^{-1}$ \\
                      & \text{} &   &  &     & $2.5$ & $6.3\times 10^{-1}$ & $2.5\times 10^{-1}$ \\ \hline
\end{tabular}}}
\end{center}
\end{table*}
In this subsection, we will first use the simulated standard siren data from Taiji alone to constrain cosmological parameters, and then briefly discuss the comparison among different space-based GW detectors. The results are given in the sixth column of Table \ref{tab:full}. In the table, we list the standard 1$\sigma$ error $\sigma(\xi)$ for every cosmological parameter $\xi$, and the three values in a cell represent different MBHB models. Here, $\sigma(\xi)$ is the absolute error, and we also use the relative error $\varepsilon(\xi)=\sigma(\xi)/\xi$ in the following discussions.

\begin{figure}[htb]
{\includegraphics[width=0.9\linewidth,angle=0]{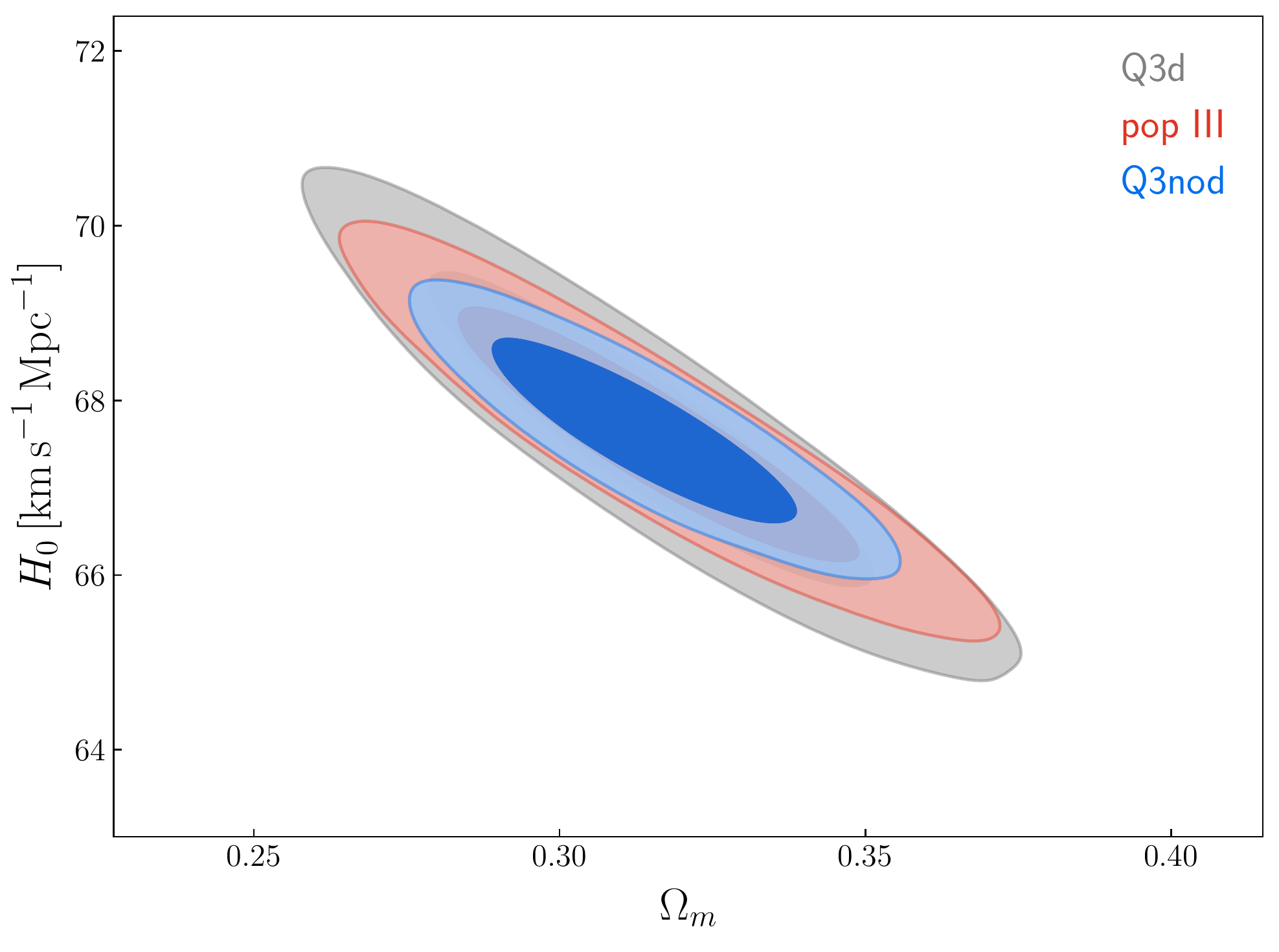}}
\caption{Two-dimensional marginalized contours (68.3\% and 95.4\% confidence level) in the $\Omega_{m}$--$H_{0}$ plane for the $\Lambda$CDM model, by using the Taiji data alone. Three MBHB models are denoted by different colors, i.e. grey (Q3d), red (pop III), and blue (Q3nod).} \label{3model}
\end{figure}

We find that the constraints of cosmological parameters based on the Q3nod MBHB model are always the best for all the cosmological models. The reason is obvious, i.e., the predicted event number in the Q3nod model is maximal among the three MBHB models. In Fig.~\ref{3model}, we show the marginalized posterior probability distribution contours in the $\Omega_{m}$--$H_{0}$ plane for the $\Lambda$CDM model as an example. Quantitatively, the data from Taiji in the Q3nod model can achieve the relative errors $\varepsilon(\Omega_{\rm m})=5.1\%$ and $\varepsilon(h)=1.0\%$ for $\Lambda$CDM, and $\varepsilon(w)=18\%$ for $w$CDM. As a contrast, the Q3d model leads to the results $\varepsilon(\Omega_{\rm m})=7.7\%$ and $\varepsilon(h)=1.8\%$ for $\Lambda$CDM, and $\varepsilon(w)=31\%$ for $w$CDM. For the CPL model, the Taiji data alone cannot constrain the cosmological parameters effectively, and so we do not discuss it here.

In the last part of this subsection, we briefly discuss the comparison of different space-based GW detectors (i.e. Taiji, LISA, and TianQin). From the results in Ref.~\cite{Wang:2019tto}, we find that the constraints on cosmological parameters from Taiji are similar with those from LISA and always tighter than those from TianQin. The reason is that the arm length of TianQin is smaller by about one order of magnitude than that of Taiji or LISA.

\subsection{Combination with CMB}

\begin{figure}[htb]
{\includegraphics[width=0.9\linewidth,angle=0]{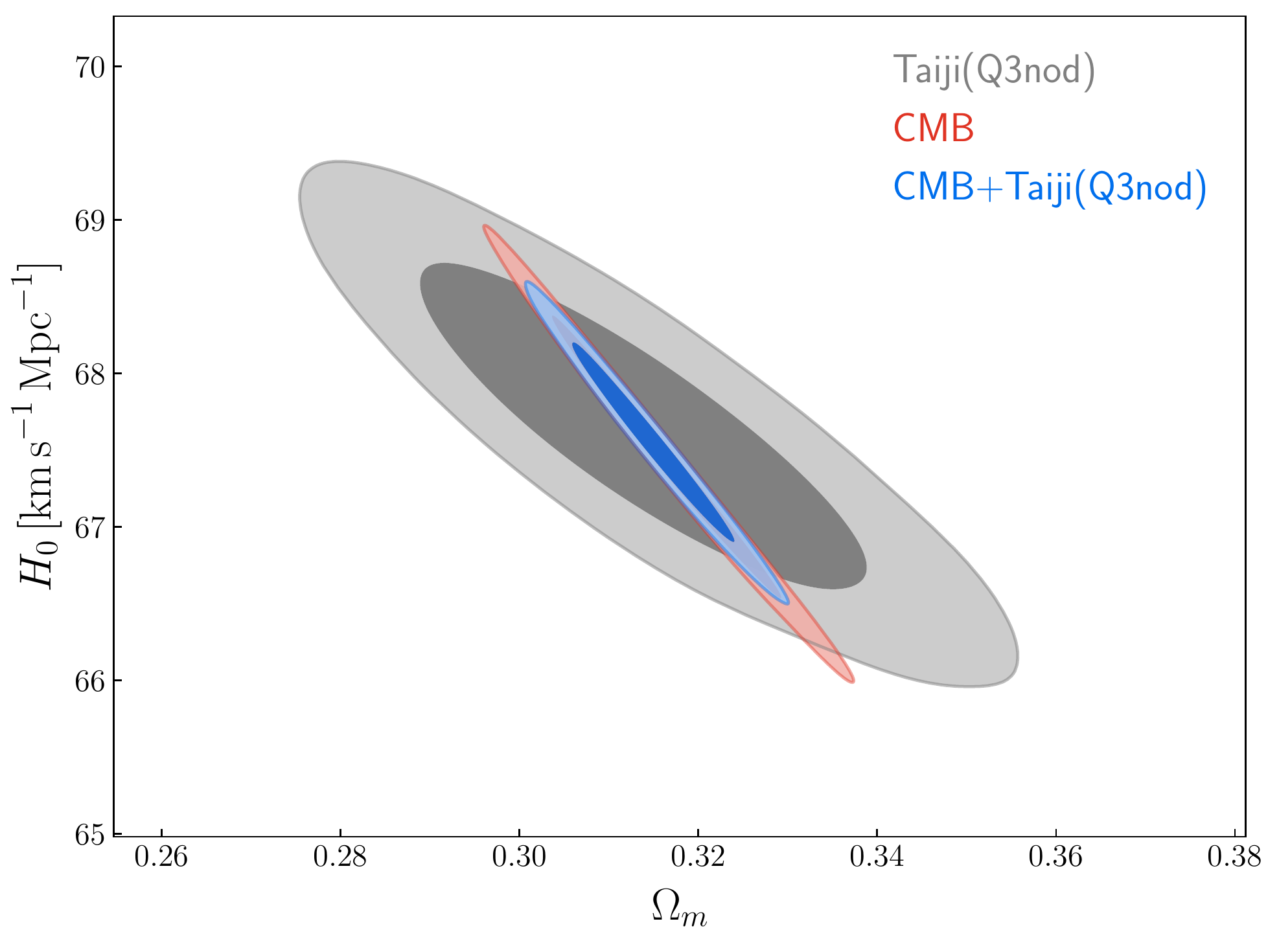}}
\caption{Two-dimensional marginalized contours (68.3\% and 95.4\% confidence level) in the $\Omega_{m}$--$H_{0}$ plane for the $\Lambda$CDM model, by using Taiji(Q3nod), CMB and CMB+Taiji(Q3nod).} \label{L-CT}
\end{figure}

\begin{figure*}[htb]
{\includegraphics[width=0.7\linewidth,angle=0]{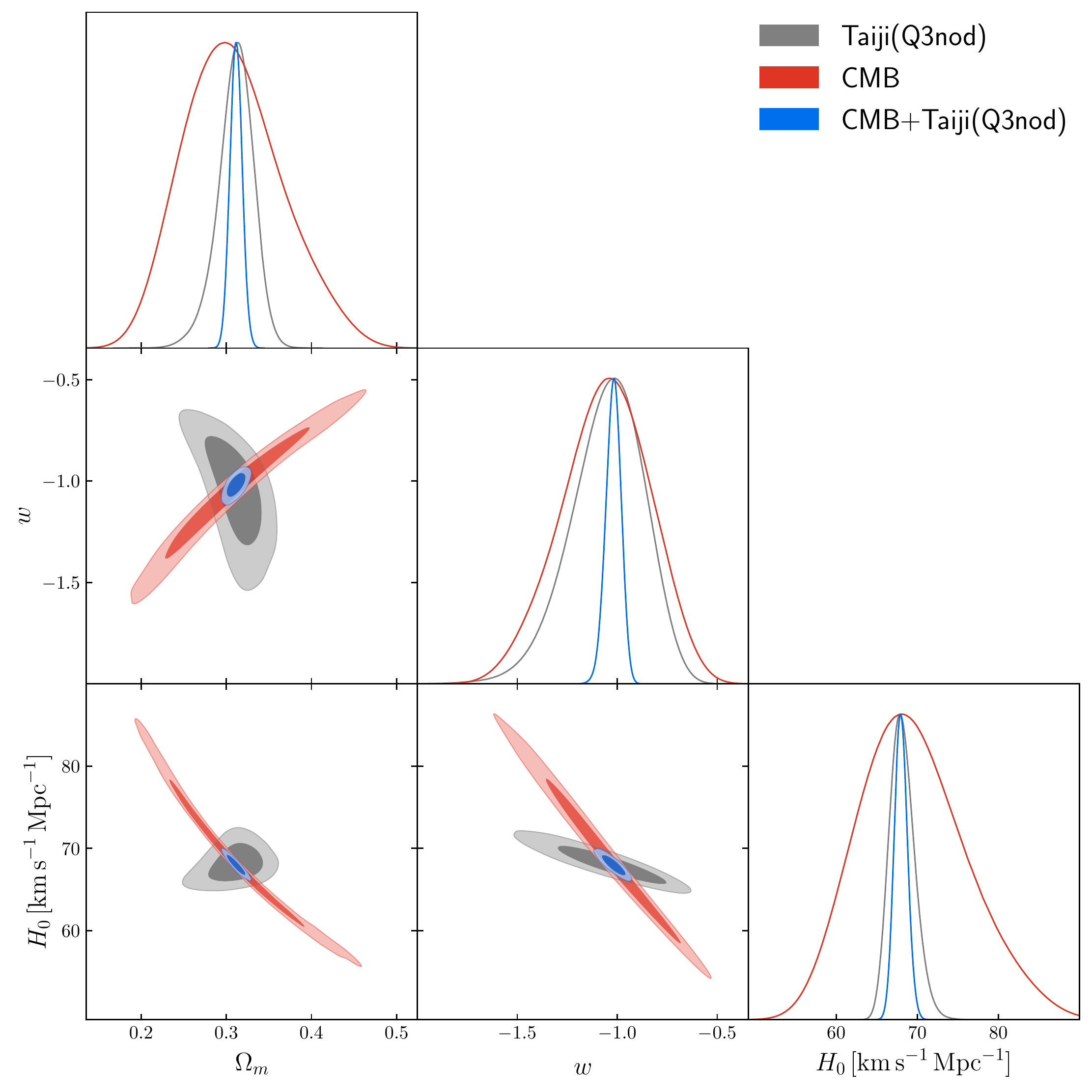}}
\caption{Constraints (68.3\% and 95.4\% confidence level) on the $w$CDM model by using Taiji(Q3nod), CMB, and CMB+Taiji(Q3nod).} \label{w-CT}
\end{figure*}

\begin{figure}[htb]
{\includegraphics[width=0.9\linewidth,angle=0]{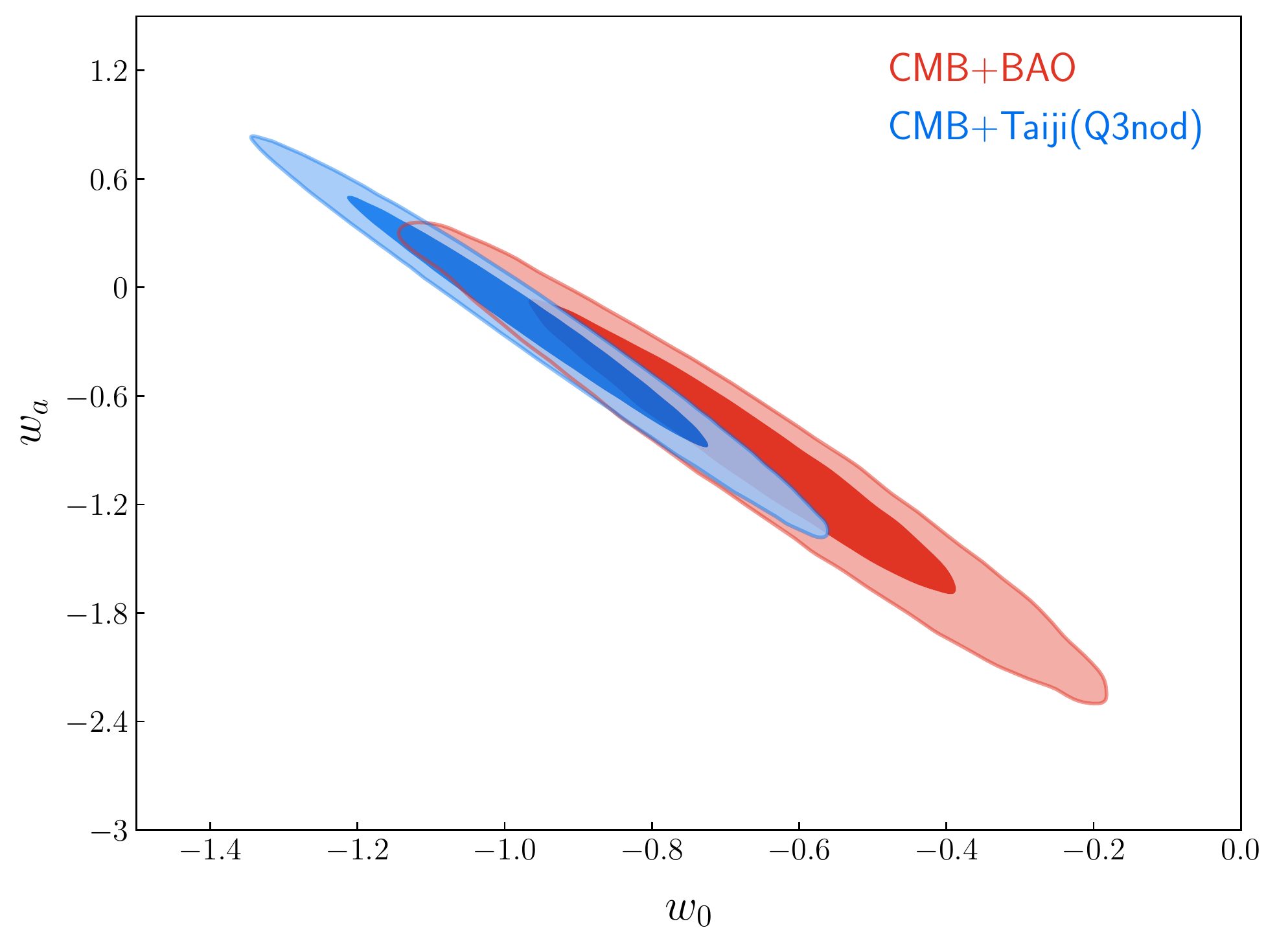}}
\caption{Two-dimensional marginalized contours (68.3\% and 95.4\% confidence level) in the $w_{0}$--$w_{a}$ plane for the CPL model, by using CMB+BAO and CMB+Taiji(Q3nod).} \label{C-CB}
\end{figure}

In this subsection, the simulated standard siren data will be combined with the CMB data to study its help in improving the cosmological parameter constraints, and the results will be compared with those from the data combination CMB+BAO. The results of the CMB, CMB+BAO, Taiji, and CMB+Taiji are given in the third, fourth, sixth, and seventh columns of Table \ref{tab:full}, respectively.

We show the contours for $\Lambda$CDM in Fig.~\ref{L-CT}, with only the best case of Taiji (i.e. the Q3nod model). It is obvious that the constraints from Taiji are weaker than those from CMB for the $\Lambda$CDM model. We can see that the combined CMB+Taiji data can still lead to evident improvement for the cosmological parameter estimation even for the $\Lambda$CDM model.
Concretely, the current CMB data combined with the simulated Taiji(Q3nod) data give the relative errors $\varepsilon(\Omega_{\rm m})=1.9\%$ and $\varepsilon(h)=0.64\%$ for the $\Lambda$CDM model, indicating that the constraints on the parameters are both improved by about 29\% compared with those using the CMB data alone.

In Fig.~\ref{w-CT}, we show the constraint results for the $w$CDM model. We can see that in this case in the $\Omega_{\rm m}$--$H_0$ plane the contours from CMB and Taiji are roughly orthogonal, and thus the parameter degeneracy generated by the CMB observation is thoroughly broken by the GW standard siren observation of Taiji. The GW observation can provide a rather good measurement for the Hubble constant $H_0$, and so the parameter degeneracy directions generated by the GW observation are different from those by the CMB observation. We know that the CMB data alone cannot tightly constrain the EoS of dark energy, $w$, which can be directly seen in Fig.~\ref{w-CT}. Actually, the GW observation from Taiji also cannot provide a tight constraint on $w$. However, due to the parameter degeneracies being broken, the combination of the two can offer a rather tight constraint on $w$. The combination of CMB+Taiji(Q3nod) gives the relative constraint errors $\varepsilon(\Omega_{\rm m})=2.3\%$, $\varepsilon(h)=1.2\%$, and $\varepsilon(w)=3.8\%$ for $w$CDM. We find that, compared to the current CMB data, the constraints on $\Omega_{\rm m}$, $h$, and $w$ are improved by 87\%, 88\%, and 83\%, respectively.

For the CPL model, neither the CMB data nor the GW data alone can provide tight constraints on $w_0$ and $w_a$. Actually, the CMB data alone can only provide a rather poor constraint on the EoS of dark energy. However, due to the degeneracy being broken by the GW data, the inclusion of the GW standard siren data from Taiji improve the constraints on $w_0$ and $w_a$ by 77\% and 80\%, respectively.

The BAO data, as a low-redshift observation, are often adopted as a complement to combine with the CMB data to constrain dark energy models. Therefore, we wish to compare the cases of CMB+BAO and CMB+Taiji (see Table~\ref{tab:full}). We find that, for the $\Lambda$CDM model, CMB+Taiji(Q3nod) can provide similar constraints as CMB+BAO. For the $w$CDM and CPL models, CMB+Taiji(Q3nod) can provide better constraints than CMB+BAO. For example, as shown in Fig.~\ref{C-CB}, the 1$\sigma$ errors on $w_0$ and $w_a$ are 0.29 and 0.77, respectively, by CMB+BAO, and 0.23 and 0.63, respectively, by CMB+Taiji(Q3nod). The results indicate that the GW standard siren data from Taiji may be an important cosmological probe to be combined with CMB in the future.

\subsection{Combination with CMB+BAO+SN}

\begin{table}[!h]
\caption{Standard 1$\sigma$ errors $\sigma$ and relative errors $\varepsilon$ on the cosmological parameters in the $\Lambda$CDM model in the ideal scenario, using CBS, Taiji, and CBS+Taiji. Here, CBS stands for CMB+BAO+SN. The three values in a cell in the columns Taiji and CBS+Taiji represent pop III, Q3d, and Q3nod, respectively, from top to bottom.}
\label{tab:LCDM}
\hspace{-0.5cm}
\begin{center}
\scalebox{0.85}{\centerline{\begin{tabular}{|c|p{2.5cm}<{\centering}|p{2.5cm}<{\centering}|p{2.5cm}<{\centering}|c|c|c}
\hline
Model & \multicolumn{3}{c|}{$\Lambda$CDM}\\
\hline
Data & CBS & Taiji & CBS+Taiji\\
\hline
  & \text{} &  $1.2\times 10^{-2}$ & $4.3\times 10^{-3}$\\
  $\sigma(\Omega_{\rm m})$& $6.0\times 10^{-3}$ &  $1.4\times 10^{-2}$ & $4.8\times 10^{-3}$\\
  & \text{} & $8.7\times 10^{-3}$ & $3.6\times 10^{-3}$ \\
\hline
  & \text{} &  $5.6\times 10^{-3}$ & $3.0\times 10^{-3}$\\
  $\sigma(h)$& $4.4\times 10^{-3}$ & $7.1\times 10^{-3}$ & $3.5\times 10^{-3}$\\
  & \text{} &  $4.0\times 10^{-3}$ & $2.5\times 10^{-3}$ \\
\hline
  & \text{} &  $3.8\times 10^{-2}$ & $1.4\times 10^{-2}$\\
  $\varepsilon(\Omega_{\rm m})$& $1.9\times 10^{-2}$ & $4.5\times 10^{-2}$ & $1.5\times 10^{-2}$\\
  & \text{} &  $2.8\times 10^{-2}$ & $1.1\times 10^{-2}$ \\
\hline
  & \text{} &  $8.2\times 10^{-3}$ & $4.4\times 10^{-3}$\\
  $\varepsilon(h)$& $6.5\times 10^{-3}$ &  $1.0\times 10^{-2}$ & $5.2\times 10^{-3}$\\
  & \text{} &  $5.9\times 10^{-3}$ & $3.7\times 10^{-3}$\\
\hline
\end{tabular}}}
\end{center}
\end{table}

\begin{figure*}[htb]
\begin{center}
\subfigure{\includegraphics[width=0.45\linewidth,angle=0]{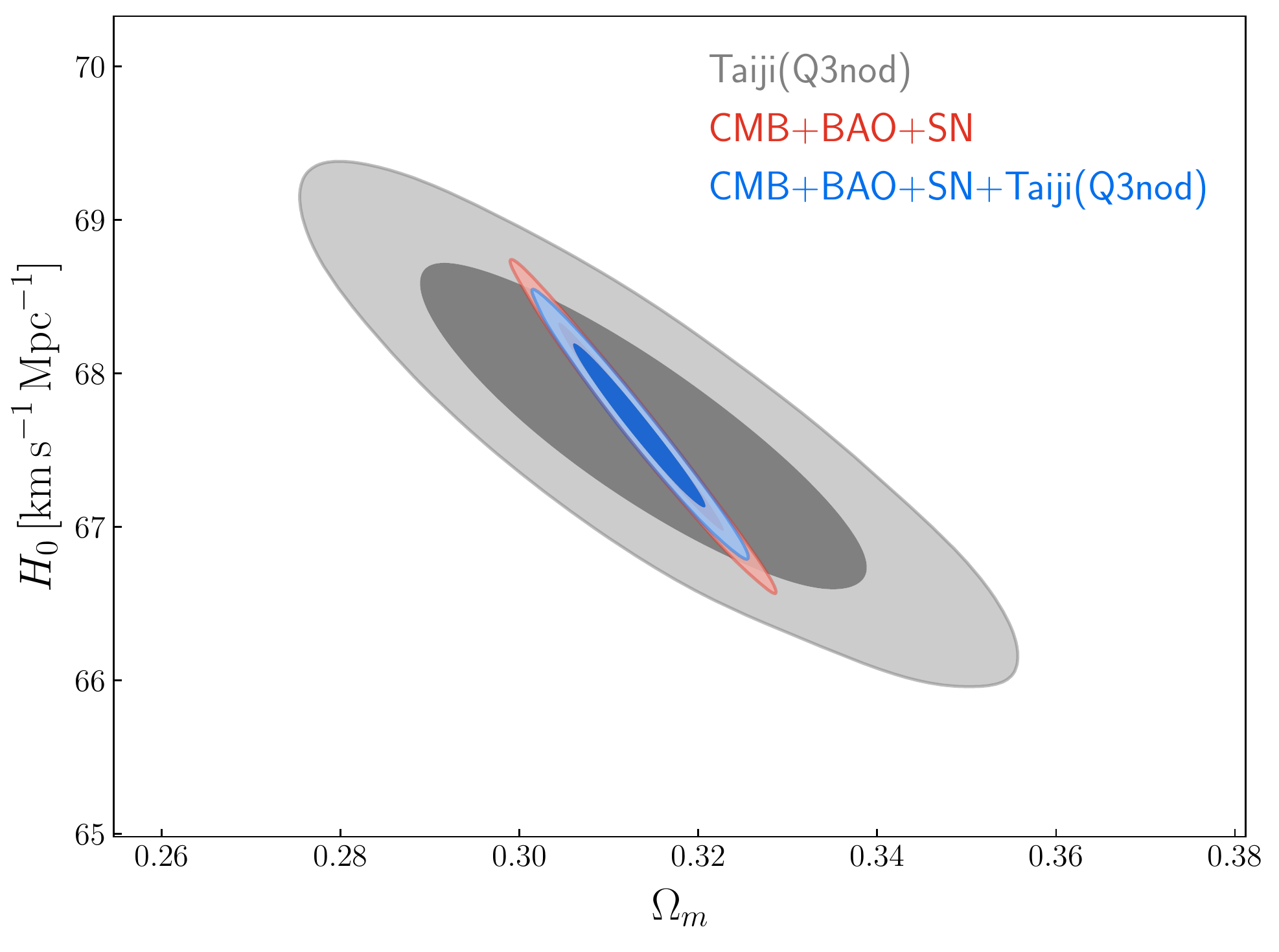}
\label{Fig.sub.1}}
\hspace*{.1cm}
\subfigure{\includegraphics[width=0.45\linewidth,angle=0]{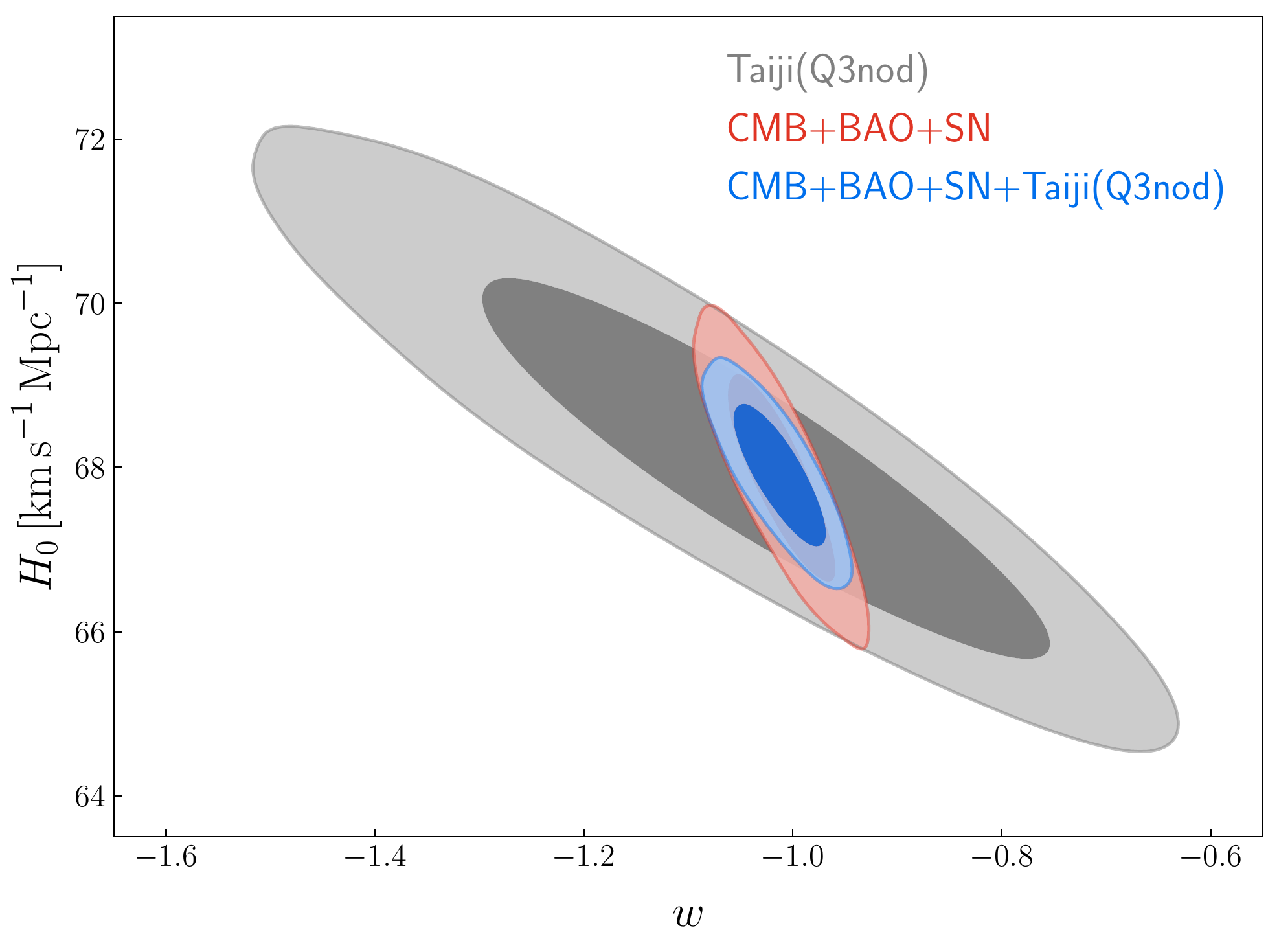}\label{Fig.sub.2}}
\end{center}
\caption{Two-dimensional marginalized contours (68.3\% and 95.4\% confidence level) in the $\Omega_{m}$--$H_{0}$ plane for the $\Lambda$CDM model (left panel) and in the $w$--$H_{0}$ plane for the $w$CDM model (right panel), by using CMB+BAO+SN, Taiji(Q3nod), and CMB+BAO+SN+Taiji(Q3nod).} \label{CBS-Taiji2}
\end{figure*}

\begin{figure}[htb]
\includegraphics[width=0.9\linewidth,angle=0]{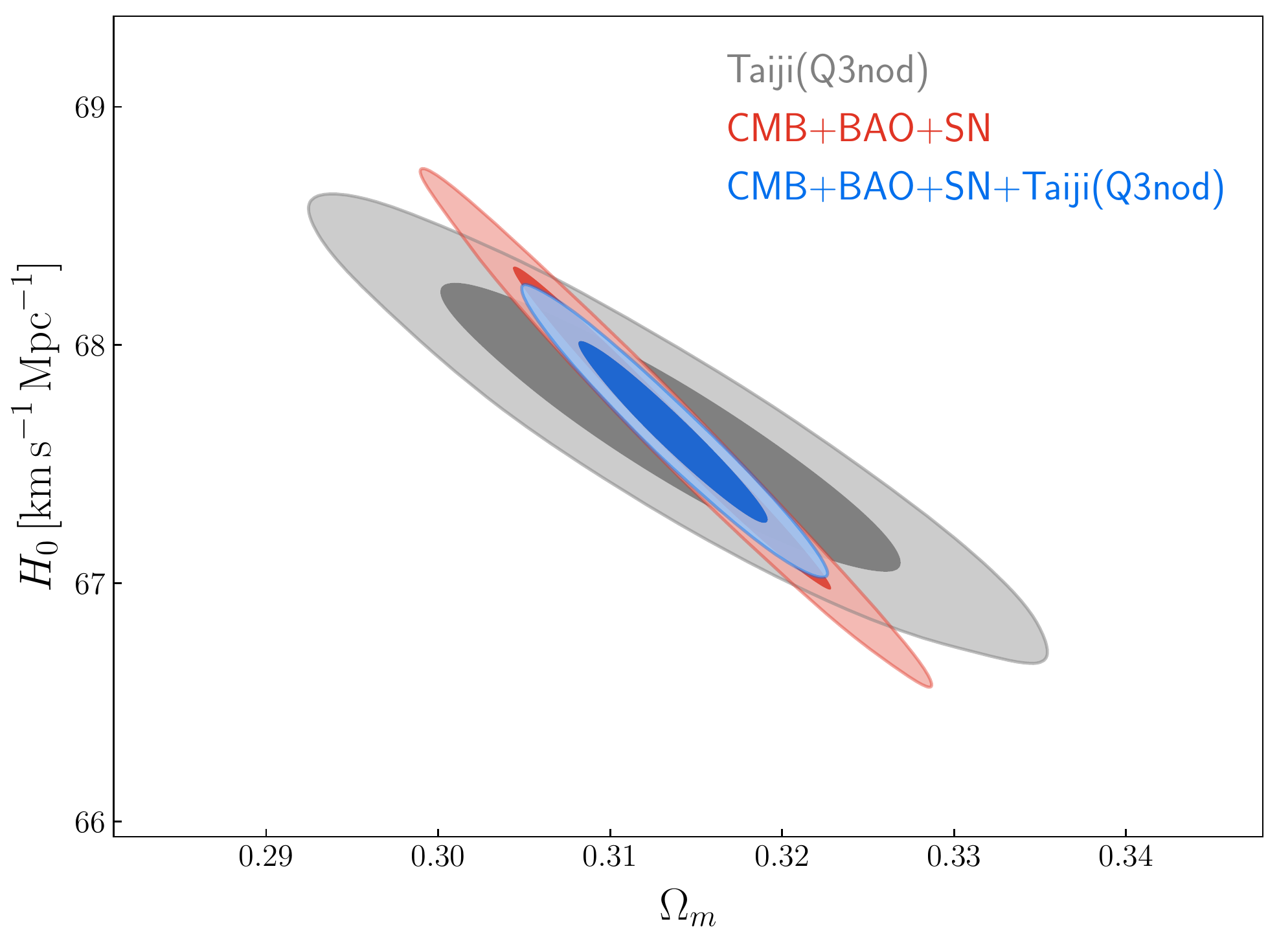}
\caption{Two-dimensional marginalized contours (68.3\% and 95.4\% confidence level) in the $\Omega_{m}$--$H_{0}$ plane for the $\Lambda$CDM model, by using CMB+BAO+SN, Taiji(Q3nod), and CMB+BAO+SN+Taiji(Q3nod). Here, for the standard siren data of Taiji, an ideal scenario is assumed.} \label{ideal}
\end{figure}

We further combine the simulated standard siren data from Taiji with the CMB+BAO+SN data. We also consider an ideal scenario to investigate the potential of Taiji in improving the cosmological parameter estimation. The results are given in Tables \ref{tab:full} and \ref{tab:LCDM}. Note that we use CBS to represent CMB+BAO+SN for convenience in the tables.

In Table \ref{tab:full}, we list the constraint results of the data combinations CMB+BAO+SN, Taiji, and CMB+BAO+SN+Taiji. The contours for the $\Lambda$CDM and $w$CDM models are shown in Fig.~\ref{CBS-Taiji2}, where the Taiji(Q3nod) data are used. The CMB+BAO+SN data can provide almost the best constraints on cosmological parameters so far, and actually the GW data alone from Taiji can only provide loose constraints on cosmological parameters (except for the Hubble constant). But owing to the fact that the parameter degeneracies can be broken by the standard sirens, the inclusion of the GW standard siren data from Taiji can still significantly improve the cosmological parameter estimation. The combination CMB+BAO+SN+Taiji(Q3nod) gives the constraint precisions $\varepsilon(\Omega_{\rm m})=1.6\%$ and $\varepsilon(h)=0.5\%$ for $\Lambda$CDM, and $\varepsilon(w)=2.9\%$ for $w$CDM. We find that, compared with the results of CMB+BAO+SN, the constraints on the parameters $\Omega_{\rm m}$, $h$, and $w$ are improved by 18\%, 20\%, and 15\%, respectively, by including the Taiji(Q3nod) data.


In fact, the error of luminosity distance is mainly from the weak-lensing and redshift measurements, especially at relatively high redshifts. Future optical/near-infrared surveys, like the Wide Field Infrared Survey Telescope (WFIRST, http://wfirst.gsfc.nasa.gov/) and the Euclid (http://sci.esa.int/web/euclid), may provide more precise measurements on the galaxy redshift and weak lensing \cite{Hemmati:2019gyg,Dore:2019pld,Chary:2019wyg}. Meanwhile, it is indicated that the weak-lensing error can be further reduced \cite{Hilbert:2010am}. Thus we consider an ideal scenario to investigate the potential of Taiji to constrain cosmological parameters, in which all the redshift errors are assumed to be ignorable and the weak-lensing error is reduced to 25\%. We forecast the cosmological parameter errors for the $\Lambda$CDM model from the Taiji data alone and from the combination CMB+BAO+SN+Taiji, with the results shown in Table~\ref{tab:LCDM}. We still choose the Q3nod model as an example, as shown in Fig.~\ref{ideal}. In the ideal scenario, the errors of cosmological parameters inferred from Taiji alone are reduced by about 50\% for all the three MBHB models,  compared with the results given in Sec.~\ref{subsec:st}. Therefore, in the future, Taiji may improve the cosmological parameter estimation in a more efficient way. Indeed, the combination of CMB+BAO+SN+Taiji(Q3nod) gives the constraint precisions $\varepsilon(\Omega_{\rm m})=1.1\%$ and $\varepsilon(h)=0.37\%$ for the $\Lambda$CDM model, and we find that compared with the case of CMB+BAO+SN the constraints are improved by about 40\% and 43\%, respectively, by including the Taiji(Q3nod) data.

\section{Conclusion} \label{sec:con}

In this work, we discuss the capability of Taiji, a China's space-based GW detection project, on improving the cosmological parameter estimation in the future by using the GW standard siren observation. In the data simulation of the GW standard sirens for Taiji, we consider three models for MBHB, i.e.,  pop III, Q3d, and Q3nod. We consider three typical dark energy cosmological models as examples to make an analysis, i.e., the $\Lambda$CDM, $w$CDM, and CPL models.

We find that, among the three MBHB models, the simulated GW data based on the Q3nod model can provide the tightest constraints on cosmological parameters.
In all the cases, the constraints from Taiji are similar with those from LISA and always tighter than those from TianQin.
Due to the fact that the standard sirens from Taiji can be used to break the parameter degeneracies generated by the CMB data, the combination of Planck CMB data and Taiji GW data can provide rather good constraints on dark energy parameters. We also find that the constraint capabilities of CMB+BAO and CMB+Taiji are actually similar, and thus the Taiji GW data have the comparable capability, compared with the BAO data, in breaking the parameter degeneracies generated by the CMB data. In addition, although the CMB+BAO+SN data can provide rather tight constraints on cosmological parameters, the inclusion of the Taiji GW data can still make significant improvements for the cosmological parameter estimation, which is shown by considering an ideal scenario in which it is assumed that the redshift error from future EM observation can be significantly reduced.

Apart from cosmological parameter estimation, there are other applications for space-based GW detectors. The low-frequency GWs can also be employed to test general relativity using MBHBs \cite{Zhao:2018vbe} and the extreme-mass-ratio inspirals \cite{Scharre:2001hn,Will:2004xi,Barack:2006pq,Niu:2019ywx}, to check inflationary scenario by stochastic background of the primordial GWs \cite{Zhou:2013tsa,Antusch:2016con,Bartolo:2016ami,Liu:2017hua,Liu:2018rrt}, and to bridge cosmology with particle physics in the framework of cosmological first-order phase transitions \cite{Apreda:2001tj,Caprini:2015zlo,Huang:2016odd,Cai:2017tmh,Wan:2018udw}. In the future, more information in the low-frequency GWs is expected to be dug out and we leave this to the future studies.

\begin{acknowledgements}
We are very grateful to Jing-Zhao Qi and Shao-Jiang Wang for fruitful discussions. This work was supported by the National Natural Science Foundation of China (Grants Nos. 11975072, 11690021, 11875102, and 11835009), the National Program for Support of Top-Notch Young Professionals, the Liaoning Revitalization Talents Program (Grant No. XLYC1905011), and the Fundamental Research Funds for the Central Universities (Grant No. N2005030).
\end{acknowledgements}



\end{document}